# Computing a Discrete Logarithm in O($n^3$)

Charles Sauerbier

December 2009

*Abstract: This paper presents a means with time complexity of at worst O($n^3$) to compute the discrete logarithm on cyclic finite groups of integers modulo p. The algorithm makes use of reduction of the problem to that of finding the concurrent zeros of two periodic functions in the real numbers. The problem is treated as an analog to a form of analog rotor-code computed cipher.*

The computation of a Discrete Logarithm[1] is a problem for which several algorithms[2] presently exist. The general problem is addressed from a variety of perspectives depending on the subject matter of interest. The Discrete Logarithm problem is also found addressed in conjunction with the more general problem of factoring.[3] However, the complexity of present solutions is not known to be polynomial.

An algorithm is presented that computes a Discrete Logarithm on cyclic finite groups of integers modulo p (i.e. $Z_p$). The complexity of the algorithm is at worst O($n^3$). No comprehensive review of existing algorithms is provided. Those interested in or otherwise unfamiliar with existing algorithms are referred to the references, and most any university text on the subject of computational complexity.

Most existing algorithms address the problem from the perspective of abstract algebra, working with various aspects of algebraic structures to improve the execution of "the naïve algorithm" of brute force factoring. Our solution addresses the problem by projection onto an arc of 360°.[4] This transforms the problem into one operating on the angular rotation of the cycle, rather than conventional notion of modular arithmetic. The projection allows us to address the exponentiation in the naïve algorithm independent of the value of p by arithmetic operations.

**Development of Algorithm:**

The Discrete Logarithm problem over cyclic finite groups has a well known definition as the solution of the equation $x^k = y$, given values for x and y, for the value of k, in some group G. We concern our self here with the cyclic finite multiplicative groups of integers modulo p.

Observing that we are dealing with cyclic groups, the problem is reducible as an analog to finding the concurrent zeros of two periodic functions, of which the periodicity of one tracks the exponentiation of "x" (i.e. has zeros as x, $x^2$, $x^3$, …, $x^n$) and the other a fixed periodicity of "y" (i.e. has zeros at y, y+p, y+2p,…, y+np). We map the group to the arc of 360° such that p→ 360, 1→ Θ = 360/p, $x \to x' = x * \Theta$, $y \to y' = y * \Theta$. The reader can readily verify that under our defined map the following hold:

$$(a * x) \to (a * x')$$
$$(b * y) \to (b * y')$$

---

[1] (Pretzel, 1996), (Weisstein)
[2] At time of writing a few algorithms were identified at http://en.wikipedia.org/wiki/Discrete_logarithm with references.
[3] (Weisstein, Prime Factorization Algorithms)
[4] The arc of 360° is an alternative cyclic finite multiplicative group, which could be shown isomorphic to any given instance of the discrete logarithm problem on cyclic finite multiplicative groups.

As consequence of the map, $x^k = y$ where $(a * x') = (b * y')$ for some a, b in Z. Where a, b exist $(a * x') = x^k$ for the least value of k and some a. Where the computation of "a" is a function of the exponentiation of "x" the result will produce "a" such that $x^k = y$.

Using the above we obtain the following algorithm[5] to compute the discrete logarithm:

```
Inputs: x, y, p
Outputs: k
--------------------
lv_theta = 360 / p
lv_x1 = x * lv_theta, lv_x2 = 0
lv_y1 = y * lv_theta, lv_y2 = lv_y1
lv_e1 = 1
k = 0

                [For i = 1 to p
                |    lv_x2 = 0
                |
                |             [ For j = 1 to x
                |    Loop 2  {     lv_x2 = lv_x2 + lv_x1
                |             [ Next
                |
                |    lv_x1 = lv_x2
Loop 1  {       |    lv_e1 = lv_e1 + 1
                |
                |             [ While (lv_x1 > 360)
                |    Loop 3  {     lv_x1 = lv_x1 - 360
                |             [ End While
                |
                |    If (lv_x1 == lv_y2)
                |       k = lv_e1
                |       Exit For
                |    End If
                [ Next

If (k != 0) output k.
```

Does the algorithm provided solve for k? Given we are dealing with a finite cyclic group we are assured that if k exists then the least value that is a solution for k exists in the interval $0 \leq k < p$. So we need only compute the power of x over the interval $0 \leq k < p$. The outer for-loop (*Loop 1*) of our algorithm iterates over the interval $0 < k < p$. The inner for-loop (*Loop 2*) computes the value $(a_i * x')$ as an exponentiation of x, which is equivalent under our projection to $x^i$. The while-loop (*Loop 3*) computes $(a_i * x')$ mod 360. The conditional statement then determines whether $((a_i * x')$ mod $360) = ((b * y')$ mod 360) which is equivalent to $x^i = y$ modulo p under the projection onto the arc of 360°. The algorithm therefore computes k.

---

[5] Algorithm performs an n-dimensional difference (recurrence) expression to iterate over the points defined by the zeros of corresponding periodic functions described by $x^i$ and $(y + (j*p))$. The difference expression performs a commonly known method of modular exponentiation with respect to $x^i$.

It is worth noting that an actual implementation of the algorithm needs to be mindful of precision, given the projection of the problem into the real number field. The algorithm is also amenable to implementation using various string representations that can provide better control over the precision and/or implementation's run time performance. The solution presented could also be implemented using radians, though the arithmetic operations can be an issue for some programming languages and computers.

The computational complexity of the nested for-loops (*Loop 1 & 2*) is worst case $O(n^2)$, as while the number of iterations is not influenced by the value of p, being determined by the value of x where x < p, the maximum value of n is p. Given that the value $(a_i * x')$ mod 360 is less than p under the projection the iteration of the while-loop (*Loop 3*), which produces a monotonic decreasing sequence, is limited by $p^2$. The computational complexity of the nested while-loop is also $O(n^3)$ in the worst case[6].

**Computation in Integers:**

The projection onto the arc of 360° provides evidence of the correctness of the algorithm for all p where dealing with a cyclic finite group on the integers modulo p. However, we can also directly compute the discrete logarithm using the same algorithm in the integer field for $Z_p$ groups.

```
Inputs: x, y, p
Outputs: k
--------------------
lv_x1 = x, lv_x2 = 0
lv_y1 = y, lv_y2 = lv_y1
lv_e1 = 1
k = 0

                    ⎡For i = 1 to p
                    │   lv_x2 = 0
                    │
                    │            ⎡ For j = 1 to x
                    │   Loop 2  ⎨    lv_x2 = lv_x2 + lv_x1
                    │            ⎣ Next
                    │
                    │   lv_x1 = lv_x2
            Loop 1 ⎨   lv_e1 = lv_e1 + 1
                    │
                    │            ⎡ While (lv_x1 > p)
                    │   Loop 3  ⎨    lv_x1 = lv_x1 - p
                    │            ⎣ End While
                    │
                    │   If (lv_x1 == lv_y2)
                    │      k = lv_e1
                    │      Exit For
                    │   End If
                    ⎣ Next

If (k != 0) output k.
```

---

[6] It is worth noting that in execution the algorithm is empirically observed to be on average $O(n^2)$.

**Conclusion:**

The method presented here computes, by an iterative algorithm that depends only on addition and subtraction, the discrete logarithm for cyclic finite groups on the integers modulo p. The algorithm obtains a computational complexity that is polynomial in the |G| and polynomial in the number of digits (bits) in |G| by addressing the exponentiation factor in the naïve algorithm. The correctness of the algorithm for all instances of cyclic finite groups on the integers modulo p is evidenced through projection of the problem onto the arc of 360°. The computational complexity of the presented algorithm has worst case $O(n^3)$.

**References:**

Pretzel, O. (1996). *Error-Correcting Codes and Finite Fields.* Oxford University Press.

Weisstein, E. W. (n.d.). *Discrete Logarithm*. Retrieved 12 10, 2009, from MathWorld - A Wolfram Web Resource: http://mathworld.wolfram.com/DiscreteLogarithm.html

Weisstein, E. W. (n.d.). *Prime Factorization Algorithms*. Retrieved 12 10, 2009, from MathWorld - A Wolfram Web Resource: http://mathworld.wolfram.com/PrimeFactorizationAlgorithms.html

**Appendix:**

A. c# code for algorithm computing in reals:

```
using System;
namespace DiscreteLogarithm
{
    public class Program
    {
        static void Main(string[] args)
        {
            System.UInt64 lv_lng_p = 373;
            System.UInt64 lv_lng_x = 13;
            System.UInt64 lv_lng_y = 158;
            System.Double lv_dbl_1 = (System.Double)360 / (System.Double)lv_lng_p;
            System.Double lv_dbl_vx1 = (System.Double)lv_lng_x * (System.Double)lv_dbl_1;
            System.Double lv_dbl_vx2 = 0;
            System.Double lv_dbl_vy1 = (System.Double)lv_lng_y * (System.Double)lv_dbl_1;
            System.Double lv_dbl_vy2 = lv_dbl_vy1;
            System.UInt64 lv_lng_v1 = 1;
            System.UInt64 k = 0;
            for (System.UInt64 i = 0; i < 360; i++)
            {
                lv_dbl_vx2 = 0;
                for (System.UInt64 j = 0; j < lv_lng_x; j++)
                {
                    lv_dbl_vx2 += lv_dbl_vx1;
                }
                lv_dbl_vx1 = lv_dbl_vx2;
                lv_lng_v1 += 1;
                while (lv_dbl_vx1 > 360)
                {
```

```csharp
                    lv_dbl_vx1 -= 360;
                }
                //  accounting for precision in implementation
                if (Math.Abs(lv_dbl_vy2 - lv_dbl_vx1) < 0.0000000001)
                {
                    k = lv_lng_v1;
                    break;
                }
            }
            Console.WriteLine(k);
            Console.WriteLine();
        }
    }
}
```

B. c# code for algorithm computing in integers:

```csharp
using System;
namespace DiscreteLogarithm
{
    public class Program
    {
        static void Main(string[] args)
        {
            System.UInt64 lv_lng_p = 17;
            System.UInt64 lv_lng_x = 6;
            System.UInt64 lv_lng_y = 12;
            System.UInt64 lv_lng_vx1 = lv_lng_x;
            System.UInt64 lv_lng_vx2 = 0;
            System.UInt64 lv_lng_vy1 = lv_lng_y;
            System.UInt64 lv_lng_vy2 = lv_lng_vy1;
            System.UInt64 lv_lng_v1 = 1;
            System.UInt64 k = 0;
            for (System.UInt64 i = 0; i < lv_lng_p; i++)
            {
                lv_lng_vx2 = 0;
                for (System.UInt64 j = 0; j < lv_lng_x; j++)
                {
                    lv_lng_vx2 += lv_lng_vx1;
                }
                lv_lng_vx1 = lv_lng_vx2;
                lv_lng_v1 += 1;
                while (lv_lng_vx1 > lv_lng_p)
                {
                    lv_lng_vx1 -= lv_lng_p;
                }
                if (lv_lng_vy2 == lv_lng_vx1)
                {
                    k = lv_lng_v1;
                    break;
                }
            }
            Console.WriteLine(k);
            Console.WriteLine();
        }
    }
}
```